\documentclass{ws-procs975x65}
\usepackage{graphicx}
\include{epsf}
\begin{document}

\title{PROSPECTS FOR NEW PHYSICS AT THE LHC}

\author{John ELLIS$^*$} %and C. D. AUTHOR}

\address{Theory Division, Physics Department, CERN,\\
CH-1211 Geneva 23, Switzerland\\
$^*$E-mail: John.Ellis@cern.ch}%\\
%{www.university\_name.edu}

%\author{A. N. AUTHOR}

%\address{Group, Laboratory, Street,\\
%City, State ZIP/Zone, Country\\
%E-mail: an\_author@laboratory.com}

\begin{center}
CERN-PH-TH/2010-074
\end{center}

\begin{abstract}
High-energy collisions at the LHC are now starting. The new physics agenda
of the LHC is reviewed, with emphasis on the hunt for the Higgs boson (or
whatever replaces it) and supersymmetry. In particular, the prospects for
discovering new physics in the 2010-2011 run are discussed.
\end{abstract}


\bodymatter

\section{Preamble}

Back in the dawn of prehistory, when I was a student, Murray Gell-Mann was an inspiration
to me. His work dominated the particle physics landscape that I entered then: strangeness, 
$V-A$ theory, the eightfold way, quarks, current algebra, the renormalization group, and so much more,
and laid the basis for the developments that have occurred since.
These fundamental contributions are now so embedded in the fabric of particle physics
that perhaps we sometimes forget to remember them and celebrate their originator with all the respect
he deserves. It is therefore a pleasure for me to participate in this meeting honouring
Murray and his achievements, and a privilege to be given the opportunity to speak here about
the next chapter in particle physics that is now unfolding.

\section{Open Questions Beyond the Standard Model}\label{sec:open}

There is a standard list of open questions beyond the Standard Model of particle physics.
What is the origin of particle masses, and are they due to a single elementary Higgs boson,
or to something else?
Why are there so many types of matter particles, 
and is the answer to this question related to the origin of matter in the Universe?
What is the nature of the cold dark matter that makes up some 80\% of the matter in the Universe?
How to unify the fundamental forces?
How to construct a quantum theory of gravity?

The LHC may be able to address all these questions. One of its main motivations has been
to solve the mass problem, and its experiments should tell us definitively whether or not
there exists a Higgs boson resembling that in the Standard Model. There is a large class of
models in which cold dark matter is composed of particles that were in thermal equilibrium
in the early Universe, in which case they should weigh $\sim 1$~TeV, and be produced
at the LHC. One example of such a theory is supersymmetry, which would also assist in the
unification of the fundamental forces. Measuring the masses of supersymmetric particles, if
they exist, would be a great way of testing predictions based on such theories.
Supersymmetry and extra dimensions are key aspects of string theory, the only promising
candidate for a consistent quantum theory of gravity, which could be tested in very novel
ways if the LHC produces microscopic black holes. What are the prospects that the LHC
might cast light on these enticing scenarios?

\section{Hunt for the Higgs}

\begin{figure}
\begin{center}
\psfig{file=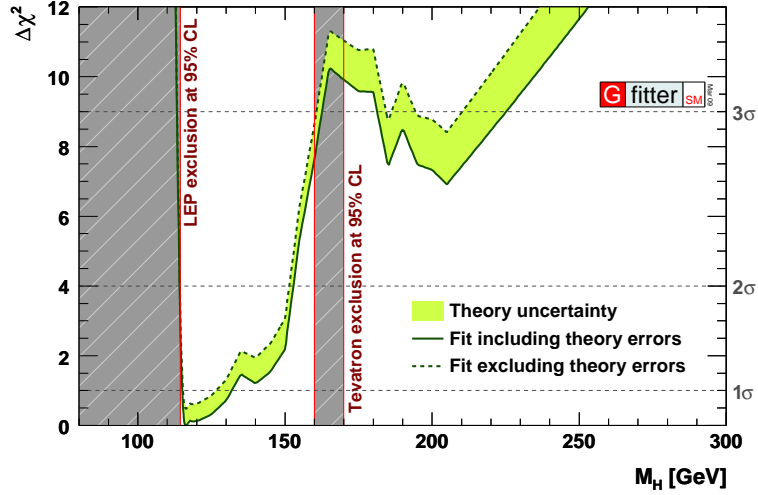,width=4in}
\end{center}
\caption{The $\chi^2$ function for the Standard Model as a function of the Higgs mass,
combining~\protect\cite{Gfitter} the precision electroweak data~\protect\cite{LEPEWWG} 
with the LEP~\protect\cite{LEP} and Tevatron~\protect\cite{Tevatron} exclusions.}
\label{fig:Gfitter}
\end{figure}

In the Standard Model, particles acquire their masses from couplings to a
universal scalar field whose associated quantum, called the Higgs boson,
has become the `Holy Grail' of particle physics. Direct searches for the
Higgs boson by experiments at the LEP accelerator established
the lower bound $m_h > 114.4$~GeV~\cite{LEP}. Precision electroweak data are
sensitive to $m_h$ through quantum corrections, and yield the preferred
range~\cite{LEPEWWG}
\begin{equation}
m_h \; = \; 87^{+35}_{-26} \; {\rm GeV}.
\label{indirect}
\end{equation}
The 95\% confidence-level upper limit on the mass of the Higgs boson is
157~GeV if only the precision electroweak data are used, or 186~GeV if
the LEP direct lower limit is included. Recently, the Tevatron experiments
CDF and D0 have excluded a range of heavier masses~\cite{Tevatron}:
\begin{equation}
162 \; {\rm GeV} \; < \; m_h \; < \; 166 \; {\rm GeV}.
\label{Tevatron}
\end{equation}
A combined fit to all the data, shown in Fig.~\ref{fig:Gfitter}, yields the asymmetric estimate~\cite{Gfitter}
\begin{equation}
m_h \; = \; 116.4^{+15.6}_{-1.3} \; {\rm GeV}
\label{all}
\end{equation}
at the 68\% confidence level.

\begin{figure}
\begin{center}
\psfig{file=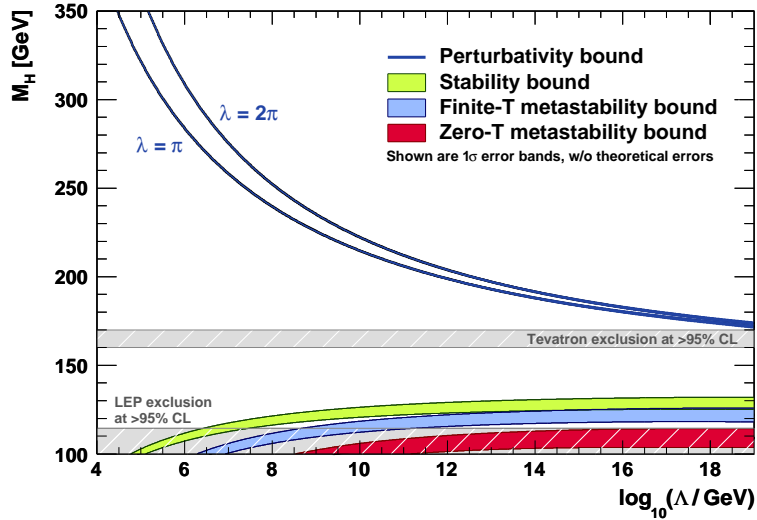,width=4in}
\end{center}
\caption{If the Standard Model Higgs boson weighs more than $\sim 180$~GeV, the
Higgs self-coupling blows up at some scale $\Lambda$ below the Planck scale, inducing
new non-perturbative physics. If it weighs less than $\sim 130$~GeV, our current
electroweak vacuum is unstable. The data summarized in Fig.~\protect\ref{fig:Gfitter}
disfavour the blow-up scenario at the 99\% confidence level~\protect\cite{EEGHR}.}
\label{fig:EEGHR1}
\end{figure}

If the Higgs mass is large, so is the Higgs quartic self-coupling $\lambda$, and renormalization-group
effects cause it to blow up at some relatively low scale $\Lambda$, as seen in
Fig.~\ref{fig:EEGHR1}, heralding the appearance of new non-perturbative physics. 
On the other hand, if $m_h$ is small, negative
renormalization by the large $t$-quark Yukawa coupling drives $\lambda < 0$,
leading to an instability in the electroweak vacuum, unless new physics such
as supersymmetry intervenes~\cite{ER}. 
Only a narrow range of $m_h \in (130, 180)$~GeV is compatible with the survival of
the Standard Model at all scales up to the Planck mass. This would be the
`maximal conceivable disaster' scenario for the LHC: a single Standard Model Higgs boson and
nothing else! The precision electroweak data favour small values of $m_h$, and the combination
with the Tevatron exclusion (\ref{Tevatron}) excludes the blow-up scenario at the 99\%
confidence level~\cite{EEGHR}. The unstable-vacuum scenario is preferred, but the `disaster' scenario
is not even disfavoured at the 1-$\sigma$ level. 

\section{The LHC Physics Haystack}

The standard list of the primary physics objectives of the LHC includes the search for
this Higgs boson (or whatever replaces it), the nature of dark matter, the primordial plasma that
filled the Universe when it was less than a microsecond old, and matter-antimatter asymmetry.
It is worth remembering that the cross sections for producing interesting heavy new
particles at the LHC are typically ${\cal O}(1$/TeV)$^2$ and possibly with additional
factors $\sim \alpha^2$, much less than the total cross section ${\cal O}(1$/$m_\pi)^2$. Typical
cross sections for producing the Higgs boson or supersymmetric particles are $\sim 10^{-12}$ 
of the total cross section: looking for them will resemble searching for a needle in 100,000
haystacks!

\section{The Search for the Higgs Boson}

\begin{figure}
\begin{center}
\psfig{file=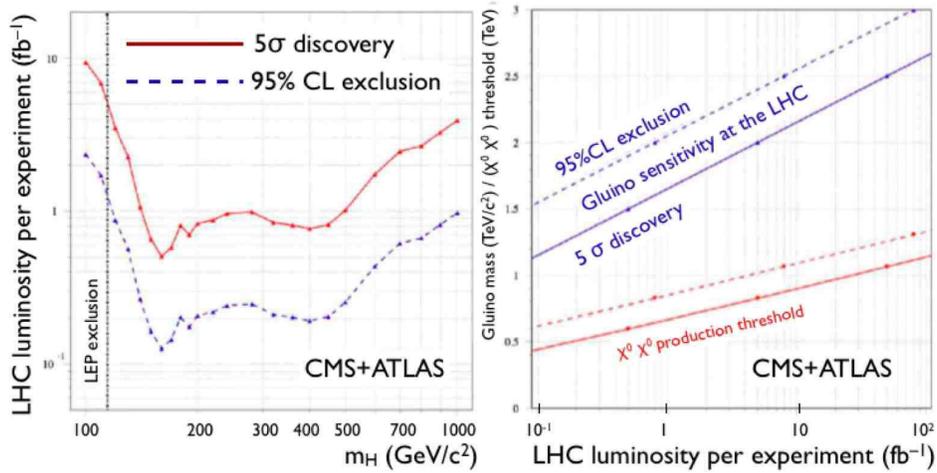,width=5in}
\end{center}
\caption{Left panel: The sensitivity of the LHC running at 14~TeV for 95\% confidence
level exclusion of the Standard Model Higgs boson (lower, dash-dotted curve) and 5-$\sigma$
discovery (upper, solid curve). Right panel: In the constrained MSSM, the mass of the gluino
(which may be excluded or discovered at the LHC, upper lines) is correlated with the threshold for
producing supersymmetric particles at a linear $e^+ e^-$ collider (lower lines)~\protect\cite{POFPA}.}
\label{fig:POFPA}
\end{figure}

For this reason, the Higgs boson will not appear as soon as the LHC produces high-energy collisions.
The left panel of
Fig.~\ref{fig:POFPA} shows the amount of luminosity required at 14~TeV in the centre of
mass either to exclude the Higgs boson
at the 95\% confidence level or to claim a 5-$\sigma$ discovery, as a function of its mass. 
We see that a couple of hundred inverse picobarns could get the LHC into the exclusion
business, while several inverse femtobarns would be needed to guarantee detection at any
mass. There is an intermediate range of masses $m_h \in (150, 500)$~GeV where detection
via $h \to WW, ZZ$ decays is relatively easy, but in the preferred low-mass range (\ref{all})
these decay modes are less important and Higgs detection becomes more difficult.

The stakes in the Higgs search are high, since it is key to many puzzles in
cosmology as well as particle physics. How is electroweak symmetry broken?
Is there such a thing as an elementary scalar field?
What is the fate of the Standard Model at high scales?
Did mass appear when the Universe was a picosecond old through an electroweak
phase transition? Did CP-violating Higgs interactions help create the matter in the Universe?
Did another elementary scalar field, the inflaton, cause (near-)exponential
expansion of the early Universe, and hence make the Universe so big and old?
The typical scale of vacuum (dark) energy in the Higgs potential is some 60 orders of
magnitude larger than the measured value: why is there so little dark energy?
Discovering the Higgs boson, or proving that it does not exist, may not answer all
these questions, but it may be our best experimental probe of them.

The most exciting Higgs scenario for the LHC may be that it is proven {\it not}
to exist. That would really shake us smug theorists out of our torpor! The best
alternative known to me would be to break electroweak symmetry by boundary
conditions in extra dimensions~\cite{Grojean}. Finding evidence for extra dimensions might be
easier to explain to a lay audience than an ill-named God particle.

\section{The Search for Supersymmetry}

There are many motivations for supersymmetry: its beauty, 
it would render the hierarchy of mass scales more natural, it predicts a light
Higgs boson, it stabilizes the electroweak vacuum, it facilitates grand unification,
and it is apparently needed for the consistency of string theory. Here I focus on
the fact that supersymmetry could provide the dark matter required by
astrophysics and cosmology~\cite{EHNOS}.

In many supersymmetric models, there is a multiplicatively-conserved
quantum number called $R$-parity, that may be represented as
$R = (-1)^{2S - L + 3B}$, where $S$ is spin, $L$ is lepton number, and
$B$ is baryon number. It is easy to verify that known particles have $R = +1$,
whereas their putative supersymmetric partners, differing in spin by 1/2,
would have $R = -1$. The conservation of $R$ parity would imply that
sparticles are produced in pairs, that heavier sparticles decay into lighter ones, and that the
lightest supersymmetric particle (LSP) is stable, because it has no legal decay mode.
Hence, it should still be around after being produced early in the Big Bang, and
could provide the needed dark matter.
Presumably, the LSP is some neutral, weakly-interacting particle, otherwise it would
have bound to ordinary matter and been detected by now. In this case, the favoured
signature for supersymmetry at colliders is missing transverse energy carried
away by the invisible dark matter particles.

\begin{figure}
\begin{center}
\psfig{file=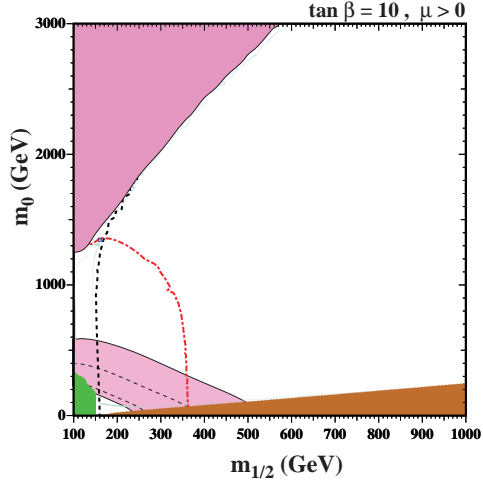,width=2.5in}
\end{center}
\caption{The $(m_{1/2}, m_0)$ plane of the CMSSM for sample values of the
other supersymmetric model parameters, showing the different theoretical,
phenomenological, experimental and cosmological constraints. There is no consistent electroweak vacuum in the dark pink shaded region at large $m_0$, the LSP would be charged in the brown shaded region
at small $m_0$, $b \to s \gamma$ excludes the green shaded region, LEP excludes the regions to the
left of the dashed black and red lines by unsuccessful chargino and Higgs searches, respectively, and $g_\mu - 2$ favours the paler shaded pink region. The LSP
would have the appropriate cosmological density in the narrow turquoise strip close to the
boundaries of the allowed region~\protect\cite{EODM}.}
\label{fig:tb10}
\end{figure}

There are important constraints on supersymmetry due to the absence
of sparticles at LEP and the Tevatron, the LEP lower limit on $m_h$
and the consistency of $b$-quark decays with the Standard Model.
Some hint of new physics at the TeV scale may be provided by the 
measurement~\cite{BNL} of the anomalous
magnetic moment of the muon, $g_\mu - 2$, that could be explained by
supersymmetry, although there are still uncertainties in the Standard Model
calculation of $g_\mu - 2$~\cite{Davier}. The measured density of dark matter,
$0.097 < \Omega_{DM} h^2 < 0.122$, provides a very tight constraint
on some combination of supersymmetric model parameters, if the LSP
provides the dark matter. The interplay of these constraints is shown in
Fig.~\ref{fig:tb10} for one particularly simple supersymmetric model with
universal supersymmetry-breaking parameters $m_{1/2}$
and $m_0$ assumed at the GUT scale, the CMSSM~\cite{EODM}.

\section{Where is Supersymmetry?}

We have recently made a global supersymmetric fit using a
frequentist approach to analyze the precision electroweak data,
the LEP Higgs mass limit, the cold dark matter density,
$b$-decay data and (optionally) $g_\mu - 2$.
We combined the likelihood functions from these different observables to
construct a global likelihood that can be used to infer preferred
regions of the supersymmetric parameter space~\cite{Master2,Master3}.

\begin{figure}[b]%
\begin{center}
\parbox{4.1in}{\epsfig{figure=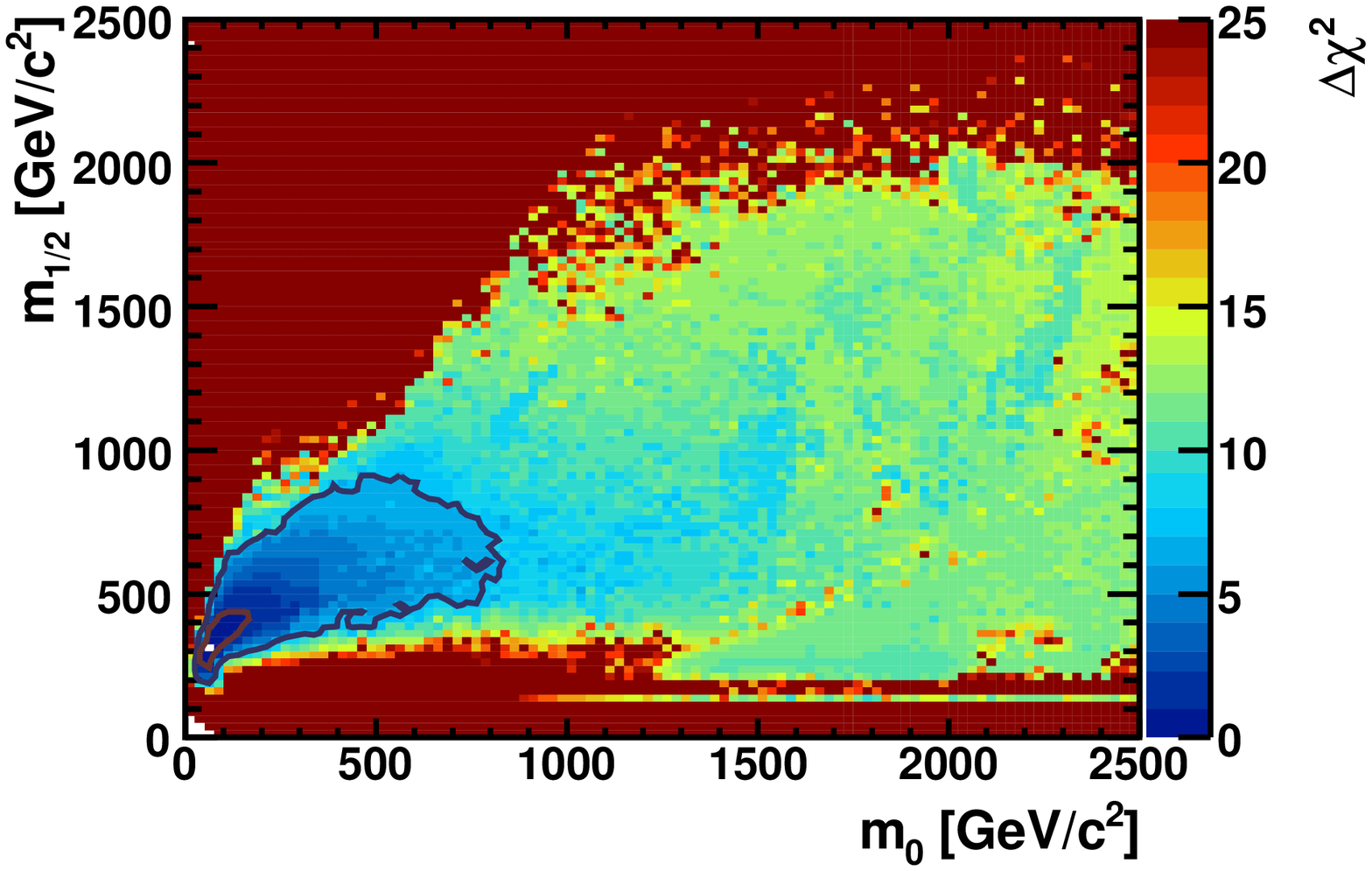,width=4.0in}}
%\hspace{-0.3cm}
% \figsubcap{a}}
 %\hspace*{4pt}
 \vspace{0.2in}
 \parbox{4.1in}{\epsfig{figure=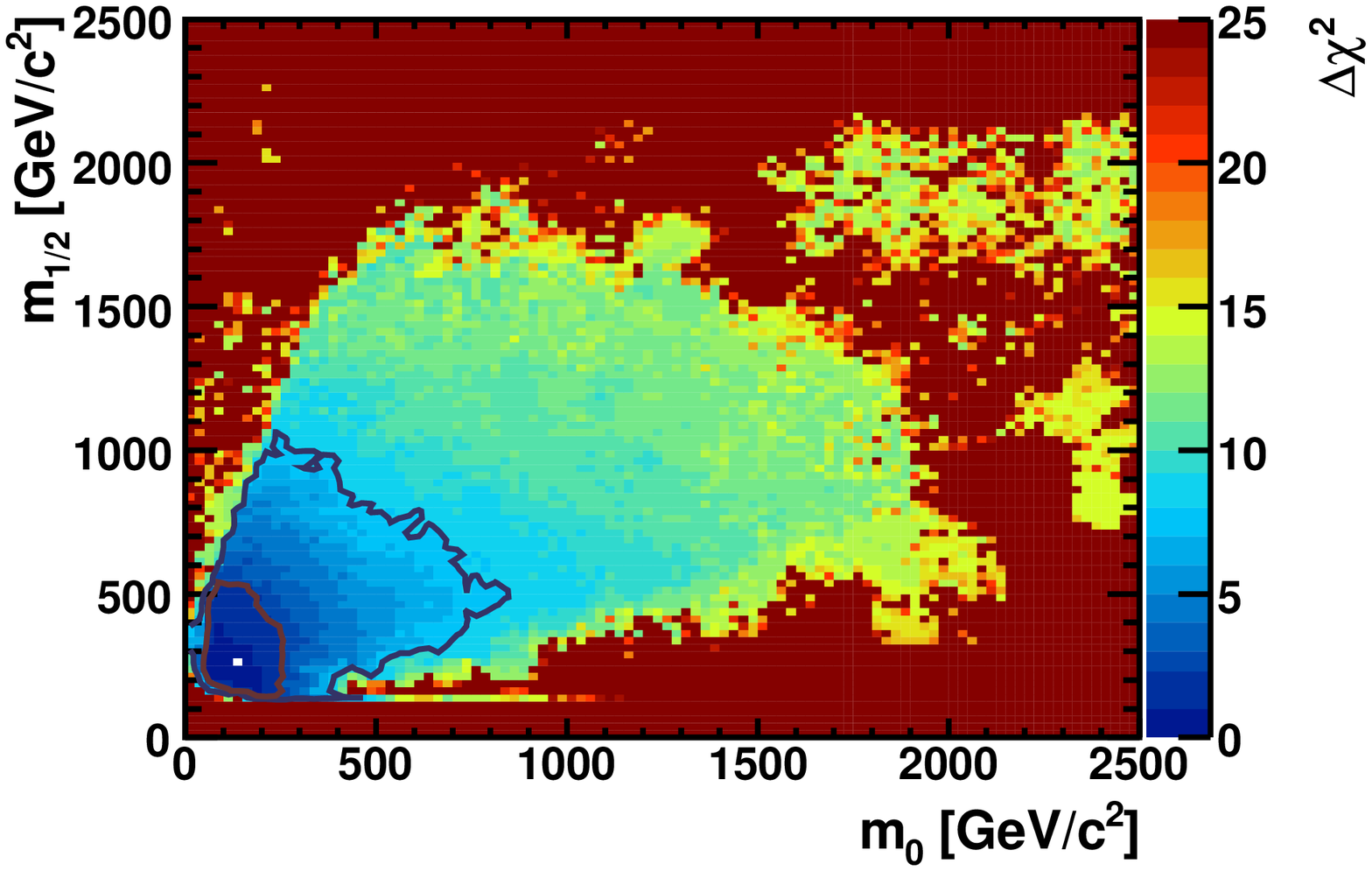,width=4.0in}}
 %\figsubcap{b}}
 \end{center}
 \caption{The preferred regions in the $(m_0, m_{1/2})$ planes of the CMSSM (upper panel)
 and the NUHM1 (lower panel), as found in a frequentist analysis~\protect\cite{Master3}.
 The best-fit points are shown as white points, and the 68\% and 95\% confidence-level
 contours are shown as solid black lines.}
\label{fig:Willis}
\end{figure}

We see in Fig.~\ref{fig:Willis} that the preferred regions of the
$(m_0, m_{1/2})$ planes in both the CMSSM and a model with common non-universal
supersymmetry-breaking contributions to the Higgs masses (NUHM1)
correspond to relatively low masses where the relic LSP density is brought
into the WMAP range by coannihilations with light sleptons, particularly the lighter stau.
The `focus-point' region at large $m_0$ is disfavoured, principally but not
exclusively by $g_\mu - 2$. If one drops this constraint, considerably larger
ranges of $m_0$ and $m_{1/2}$ would be allowed, though small values are
slightly preferred by other data~\cite{Master2,Master3}.

\begin{figure}[b]%
\begin{center}
\parbox{3.6in}{\epsfig{figure=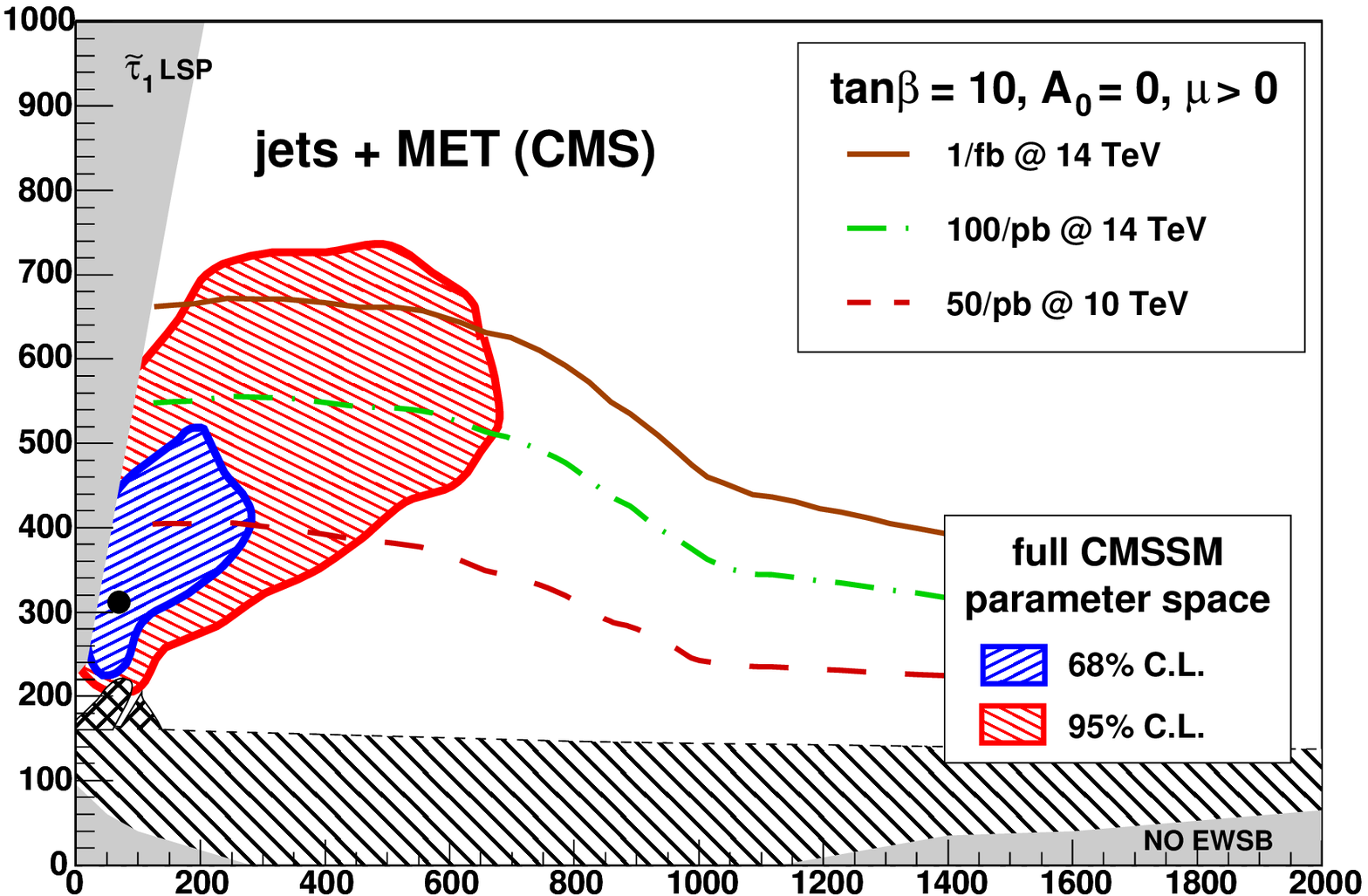,width=3.5in}}
% \figsubcap{a}}
 %\hspace*{4pt}
 \end{center}
 \vspace{0.1in}
 \begin{center}
 \parbox{3.6in}{\epsfig{figure=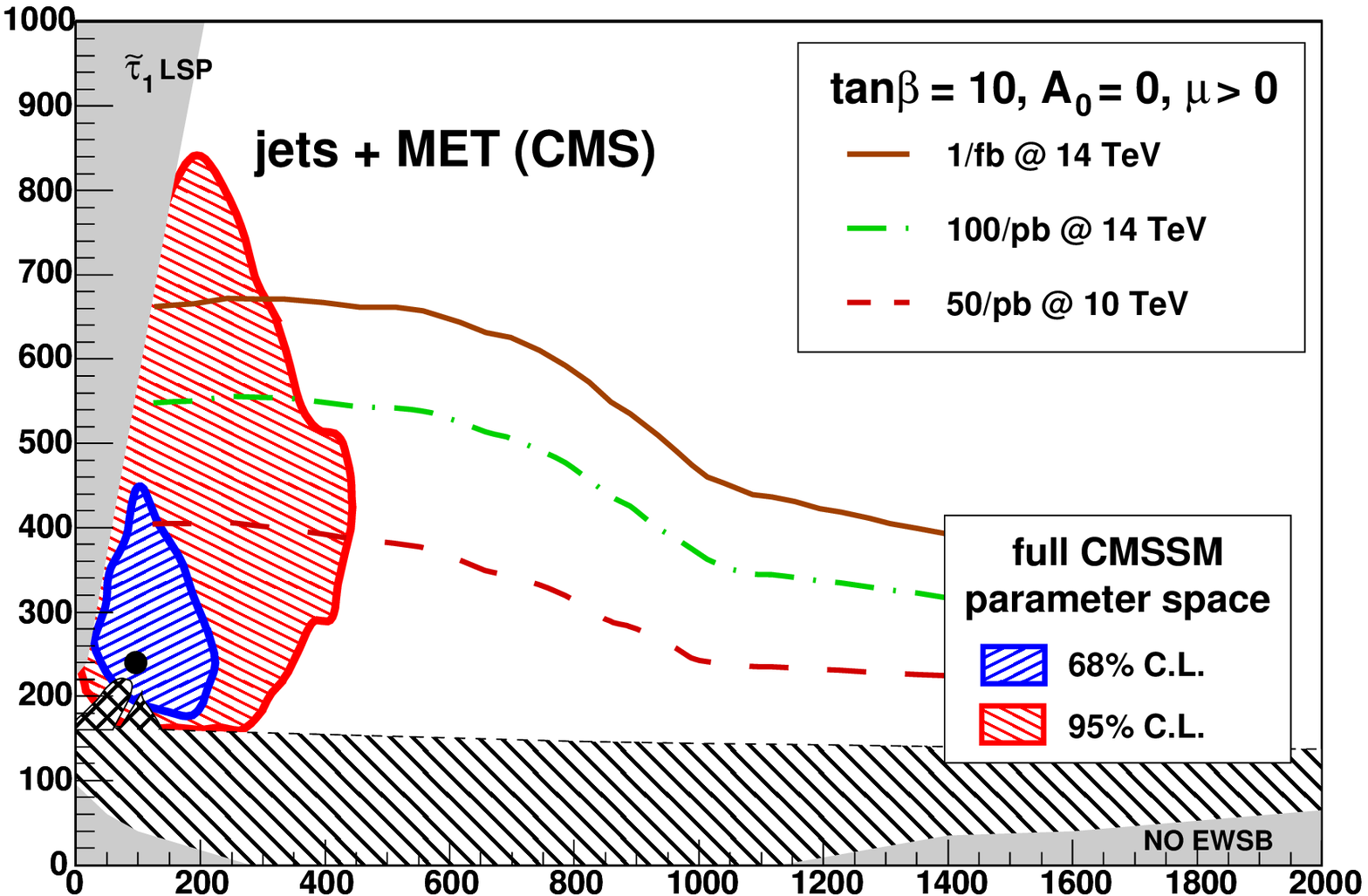,width=3.5in}}
 %\figsubcap{b}}
 \end{center}
 \caption{The preferred regions in the $(m_0, m_{1/2})$ planes of the CMSSM (upper panel)
 and the NUHM1 (lower panel), compared with the estimated discovery sensitivity
 of the LHC with different amounts of luminosity and centre-of-mass energy~\protect\cite{Master2}.}
\label{fig:discovery}
\end{figure}

Fig.~\ref{fig:discovery} compares the preferred regions of these $(m_0, m_{1/2})$ planes
with the 5-$\sigma$ discovery reach of the LHC with given amounts of integrated luminosity
at certain centre-of-mass energies. The 5-$\sigma$ discovery reach of the LHC with 1/fb of luminosity at
7~TeV in the centre of mass would probably include all the 68\% confidence-level regions
in Fig.~\ref{fig:discovery}.

If supersymmetric particles are this light, they would contribute significantly to the electroweak
radiative corrections that test the Standard Model at the quantum level. Using low-energy
precision measurements as input, one may use radiative corrections
to predict high-energy observables such as $m_t$, $M_W$ and $m_h$. Within the
Standard Model, the predictions of $m_t$ and $M_W$ have been very
successful, and one would hope that their success would not be undermined in
supersymmetric extensions of the Standard Model. On the one hand, supersymmetry
predicts a restricted range for $m_h$, tending to reduce an uncertainty in the
calculations of the radiative corrections. On the other hand, light sparticles could
themselves contribute significantly to the radiative corrections.

\begin{figure}
\begin{center}
\psfig{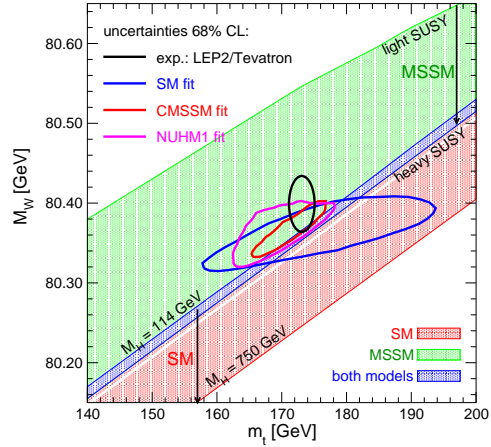}
\end{center}
\caption{Predictions for $m_t$ and $M_W$ on the basis of low-energy precision
data in the Standard Model (blue curve), the CMSSM (red curve) and the NUHM1 (violet curve)
compared with the experimental measurements at LEP and the 
Tevatron (black ellipse)~\protect\cite{Master3.5}.}
\label{fig:mtMW}
\end{figure}

As seen in Fig.~\ref{fig:mtMW}, the net effect of supersymmetry in the context of
the global fits to the CMSSM and NUHM1 introduced above is to {\it reduce}
the predicted ranges of $m_t$ and $M_W$~\cite{Master3.5}. The predictions are in both cases in
good agreement with the measured values: indeed, the supersymmetric predictions
for $M_W$ are even slightly better than those in the Standard Model.

\section{Extra Dimensions?}

Another fashionable scenario for physics beyond the Standard Model is the
appearance of large extra dimensions. Their possible existence was suggested decades
ago, and given extra momentum by string theory, which appears to require their
appearance at some distance scale. However, our current understanding of string theory
does not offer any firm guidance as to their possible sizes. It has been suggested that,
if they are large they might solve or at least alleviate the mass hierarchy problem.
Perhaps I am showing my age, but I find large extra dimensions less attractive than
supersymmetry. Nevertheless, the final arbiters will be experiments, particularly at
the LHC.

They may find Kaluza-Klein excitations of Standard Model particles, or they may discover
missing energy leaking into extra dimensions, or they may discover that gravity
becomes strong at the TeV scale and LHC collisions produce microscopic
black holes. This last speculation has sparked concerns that have no scientific basis.
The same theories that predict microscopic black holes also predict that they decay via
Hawking radiation, with interesting grey-body factors that would be
fascinating tests of string theory or any other quantum theory of gravity. 
Even if they were stable, the continued existence of the Earth and
other celestial objects tells us to fear nothing from the LHC~\cite{LSAG}. Do I hear the sound of
other axes being ground?

\section{The Restart of the LHC}

Following the first start of the LHC on September 10th, 2008 and the electrical fault nine days
later that laid the LHC low, there was jubilation on November 20th, 2009 when the LHC was
restarted. This was redoubled 3 days later when the first collisions were observed at 900~GeV in the
centre of mass. Soon afterwards, the beams were successfully ramped up to 1.18~TeV and
the highest-energy human-made collisions were seen.

Remarkably quickly, the LHC experiments were able to reconstruct the decays of
known particles and remeasure their masses. Two-photon decays of the $\pi^0$ and $\eta$
were quickly seen, but seeing the Higgs will take a while, as discussed above: no
Higgs yet! Both ATLAS
and CMS have seen multi-jet events, and have shown distributions of the
missing transverse energy. So far, these agree only too well with Monte Carlo simulations:
no supersymmetry yet! Multi-jet events occur at rates compatible with QCD: no
Hawking-decaying black holes yet!

\section{What will the Future Bring?}

\begin{figure}[t]%
\begin{center}
\epsfig{file= 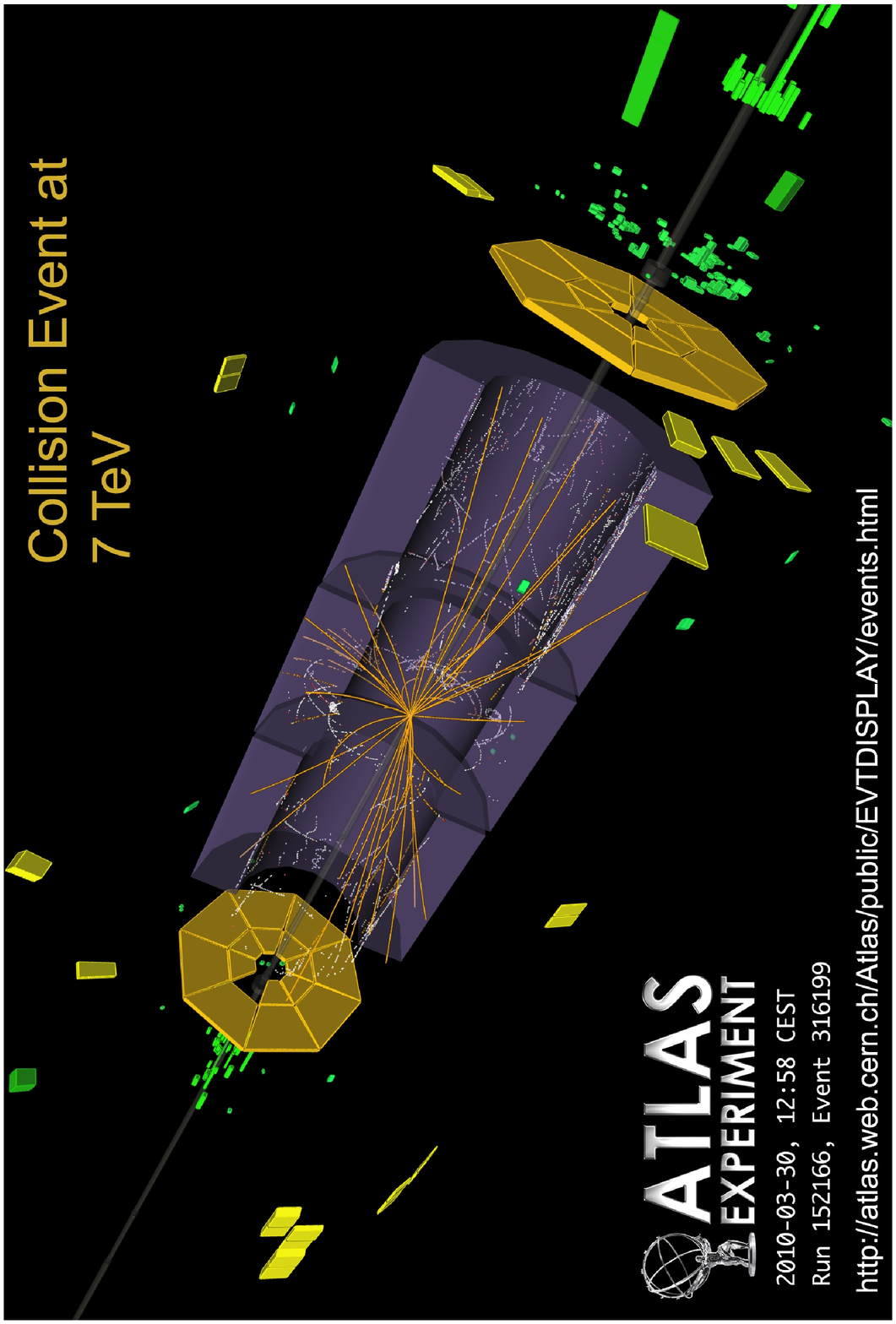, height=4.0in,angle=270}
\end{center}
\vspace{0.05in}
\begin{center}
\epsfig{file= 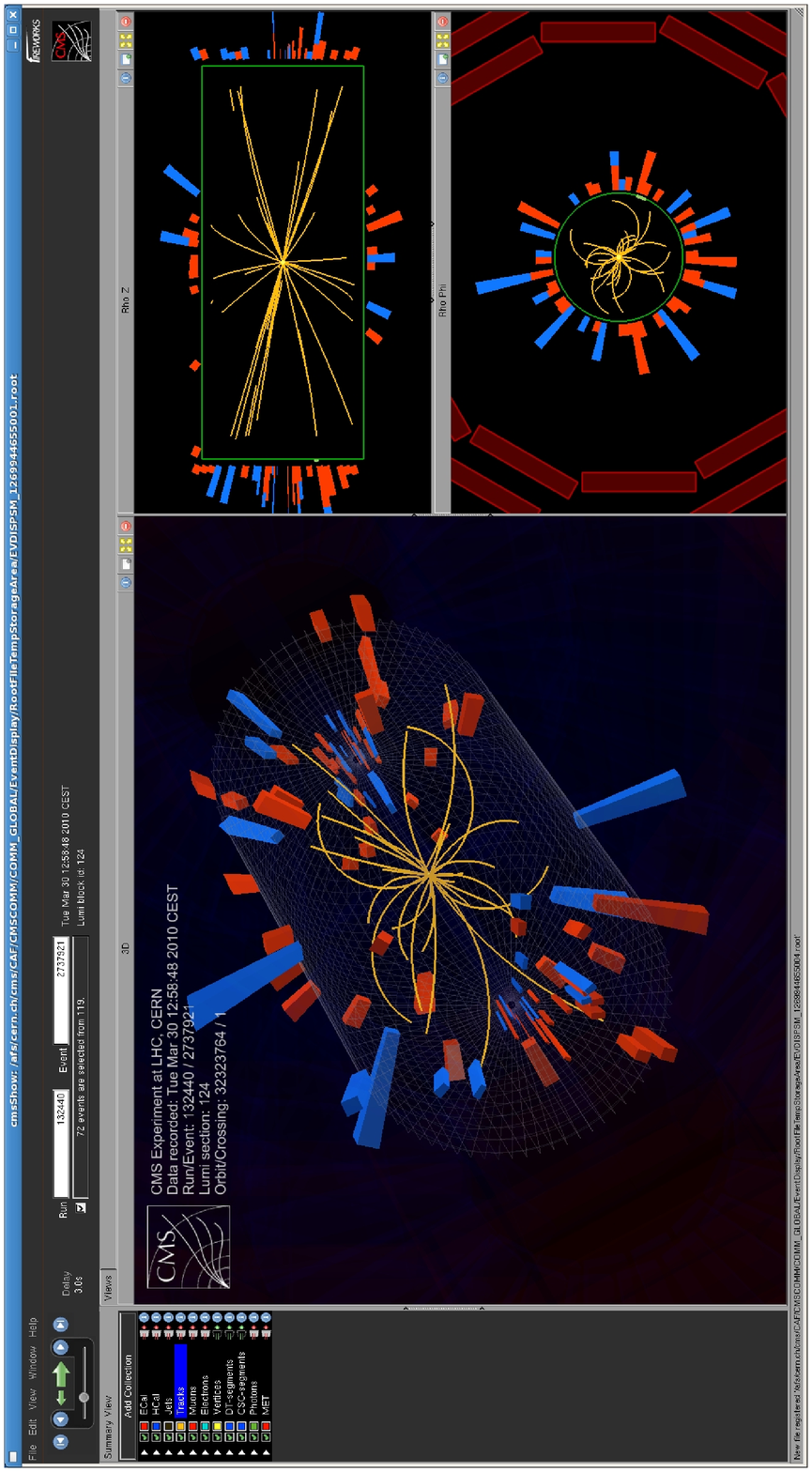, height=4.0in,angle=270}
\end{center}
 \caption{First 7-TeV collisions from ATLAS (upper panel)~\protect\cite{ATLAS} 
 and CMS (lower panel)~\protect\cite{CMS}.}
\label{fig:first}
\end{figure}

\begin{figure}[t]%
%\begin{center}
\vspace{1.5in}
\parbox{2.5in}{\epsfig{figure=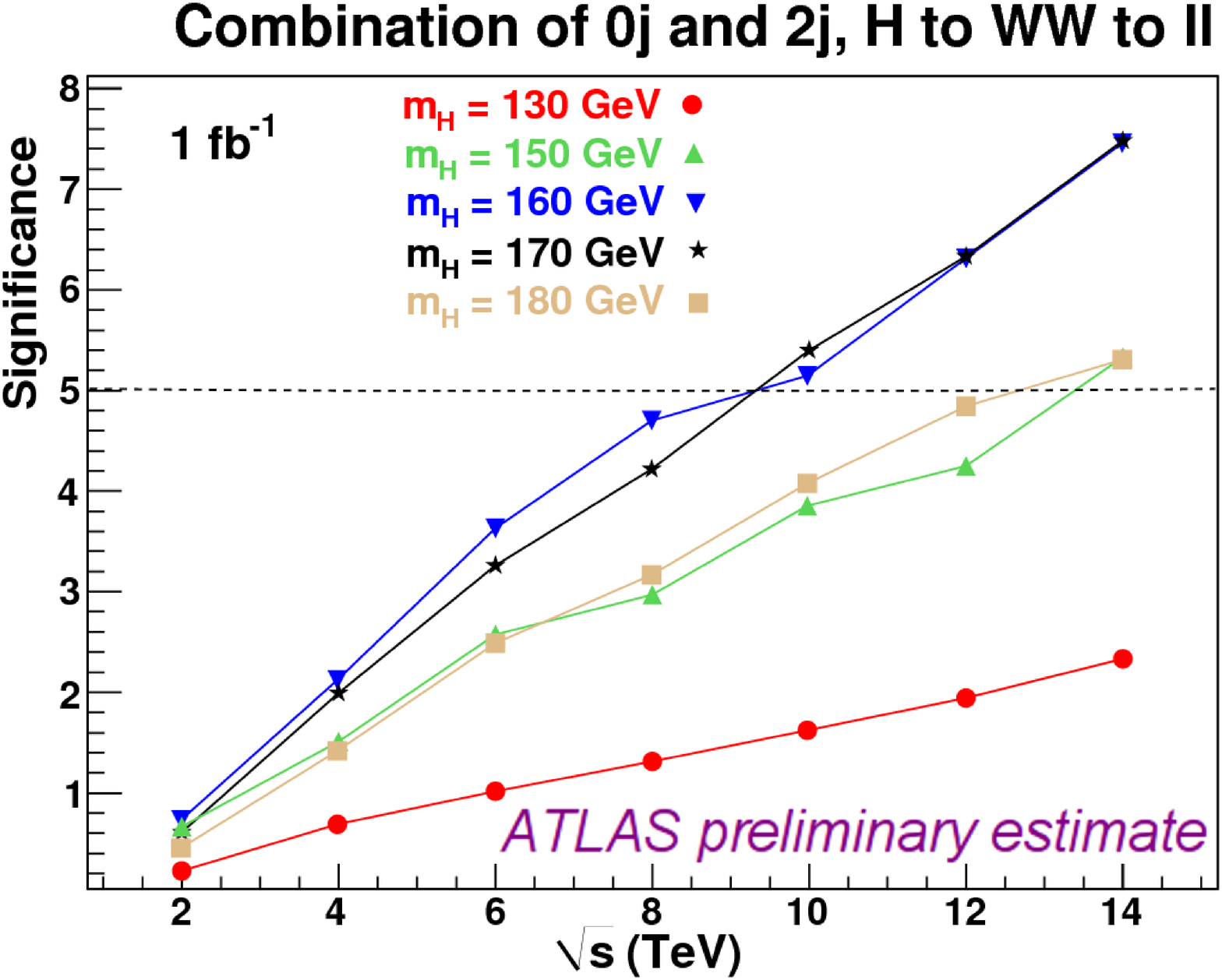,width=2.4in}}
% \figsubcap{a}}
 \hspace*{4pt}
 \parbox{2.3in}{\epsfig{figure=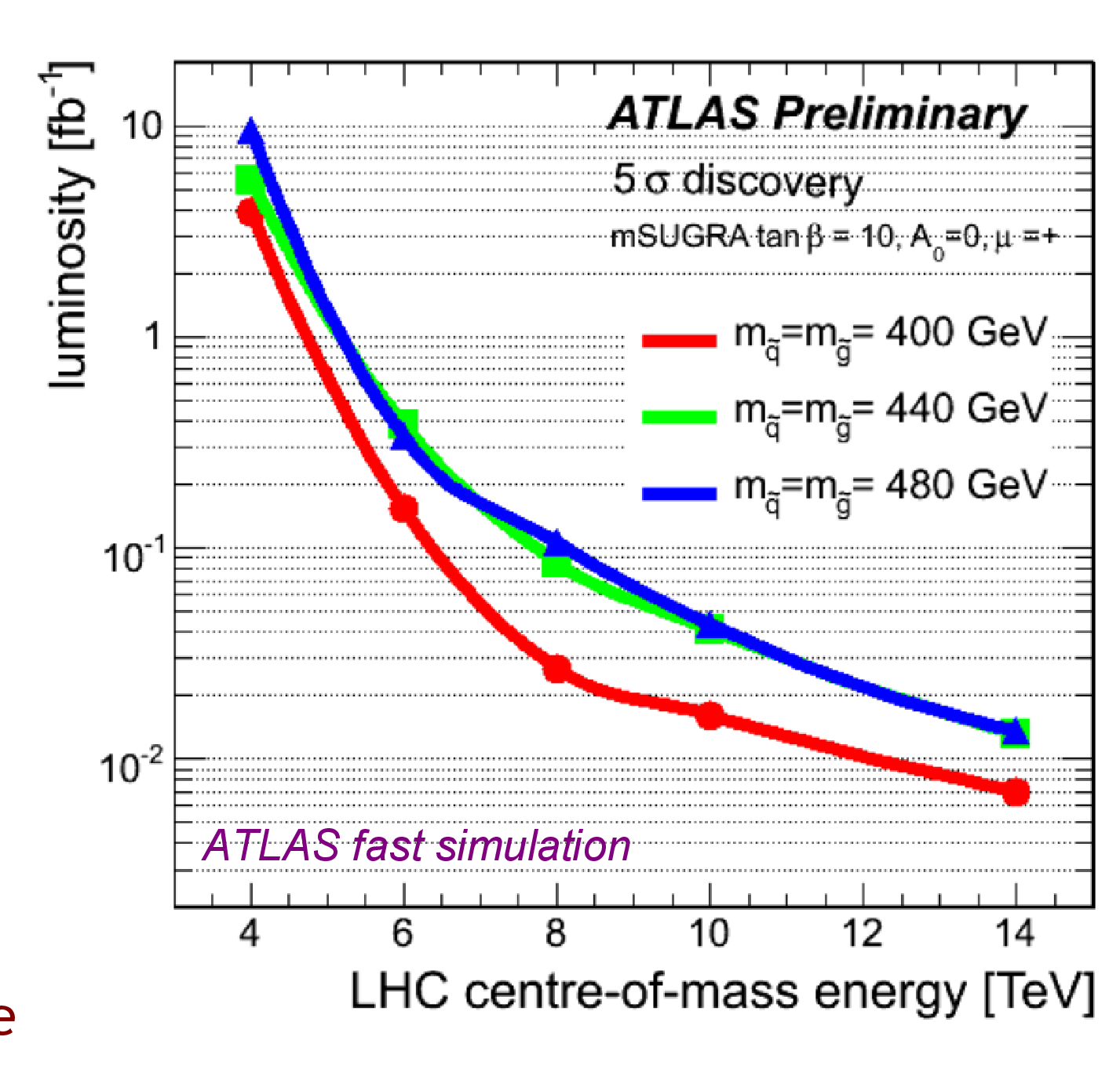,width=2.2in}}
 %\figsubcap{b}}
 %\end{center}
 %\vspace{-0.5in}
 \caption{Left panel: The ATLAS signal sensitivity for a Standard Model
 Higgs boson with 1/fb integrated luminosity at various centre-of-mass energies. Right panel: 
 The sensitivity of ATLAS for 5-$\sigma$ discovery of supersymmetric
 particles at reduced centre-of-mass energies~\protect\cite{ATLASphys}.}
\label{fig:7TeV}
\end{figure}

While I was writing up this talk, the LHC produced its first collisions at 3.5~TeV per beam
on March 30th, 2010, and Fig.~\ref{fig:first} shows
some of the first events observed by the ATLAS and CMS experiments.
The default LHC operating scenario is to collide at 3.5~TeV per beam until the
end of 2011, aiming to accumulate 1/fb of integrated luminosity. Depending on the
LHC running experience during this period, the LHC beam energy may be
increased slightly during this period. As illustrated in Fig.~\ref{fig:7TeV},
this should enable the LHC to at least
equal the Tevatron sensitivity to an intermediate-mass Higgs boson weighing
between 150 and 180~GeV. Moreover, it would give the LHC a reach for
new physics such as supersymmetry that would extend beyond the Tevatron.
Indeed, by discovering or excluding the gluino,
this first physics run of the LHC may already be able to tell us whether supersymmetric
are light enough to be produced with a 500~GeV linear $e^+ e^-$ collider~\cite{POFPA}.

A long shutdown is then planned for
consolidation of the LHC and its injectors, including the LHC magnet interconnects
and training of the dipole magnets, making it possible to run the LHC at or close to its
design energy of 7~TeV per beam.
Further in the future, there will be at least major upgrade of the LHC,
incorporating Linac4 and new interaction-region insertions. The scope of
possible further upgrades (an SPL? a higher extraction energy
for the PS booster? replacement of the PS? new collision insertions? crab cavities?)
are still under discussion, and will be decided only in the light of operational
experience with the LHC.

\section{A Conversation with Mrs Thatcher}

Back in 1982, while she was Prime Minister of the UK, Mrs. Thatcher visited CERN
and was introduced to British physicists. When she was told I was a theoretical
physicist, she asked me ``What do you do?" I explained that my job was to think
of things for the experiments to look for, and hope they find something different.
``Wouldn't it be better if they found what you predicted?" asked Mrs. T., who always
liked {\it her} ideas to be vindicated. My response was that, in that case, we would
not be learning anything really new. Likewise, I sincerely hope that the LHC will
be remembered in the history of physics for something {\it not} described in this talk.

\section*{Acknowledgments}

It is a pleasure to thank Harald Fritzsch for his kind invitation to this
enjoyable event, and K. K. Phua and his team for their generous hospitality.
Above all, it is a pleasure to thank Murray for his inspiration over the years.

\end{document}